\documentclass{spie}
\usepackage{amsfonts}
\usepackage{amsmath}
\usepackage{graphicx}

\newtheorem{conjecture}[theorem]{Conjecture}

\input{tcilatex}
\begin{document}

\title{The unified theory of chirped-pulse oscillators}
\author{Vladimir L. Kalashnikov\supit{a} \\
\supit{a}Institut f\"{u}r Photonik, TU Wien, Gusshausstr. 27/387, A-1040
Vienna, Austria }
\authorinfo{Further author information:\\
E-mail: kalashnikov@tuwien.ac.at, Telephone: +43-1-58801-38743}
\maketitle

\begin{abstract}
A completely analytical theory of chirped-pulse oscillators is presented.
The theory is based on an approximate integration of the generalized
nonlinear complex Ginzburg-Landau equation. The obtained parametric space of
a chirped-pulse oscillator allows easy tracing the characteristics of both
solid-state and fiber oscillators operating in the positive dispersion
regime.
\end{abstract}


\section{Introduction}

In the last decade, femtosecond pulse technology has evolved rapidly and
allowed achieving a few-optical-cycle pulse generation directly from an
oscillator \cite{krausz}. Applications of such pulses range from medicine
and micro-machining to fundamental physics of light-matter interaction at
unprecedented intensity level and time scale \cite%
{mourou,spielmann,agostini,krausz2}. High-energy laser oscillators nowadays
have reached the intensity level of the order of 10$^{14}$ \ W/cm$^{2}$,
which allows high-intensity experiments such as direct gas ionization \cite%
{morgner}. To achieve these regimes, about- and over-microjoule pulse
energies are required. Such energy frontiers have become achievable due to
the chirped pulse amplification outside an oscillator \cite{mourou}.
However, the amplifier technology is i) complex, ii) expensive, iii)
accessible pulse repetition rates lie within the kHz range, and iv) noise
amplification is unavoidable.

It is desirable to find a road to the direct over-microjoule femtosecond
pulse generation at the MHz pulse repetition rates without an external
amplification. To date, a promising approach has been proposed. It is based
on a considerable decrease of the oscillator repetition rate \cite%
{fujimoto,krausz3}. The catch is that a long-cavity oscillator suffers from
strong instabilities caused by enhanced nonlinear effects owing to increase
of the pulse peak power $P\left( 0\right) $. The leverage is to stretch a
pulse and, thereby, to decrease its peak power below the instability
threshold. Recently, a critical milestones, demonstrating the feasibility of
this approach, has been achieved for the Ti:sapphire oscillators operating
both in the negative- (NDR) \cite{fujimoto2,fujimoto3} and
positive-dispersion regimes (PDR) \cite{fernandez,naumov}, the near-infrared
Yb:YAG thin-disk oscillators operating in the NDR \cite%
{keller,neuhaus,morgner2}, and the fiber oscillators operating in the
all-normal dispersion (ANDi) regime (that is the PDR by definition) \cite%
{wise}.

The fundamental difference between the NDR and the PDR is that, in the first
one, the Schr\"{o}dinger soliton develops \cite{kaertner}. Since the soliton
peak power $P\left( 0\right) $ has to be lower than some threshold value $%
P_{th}$ in order to avoid the soliton destabilization, the maximum reachable
energy can be estimated as $E=2P_{th}T$ ($T$ is the soliton width). That is
the energy scaling requires the pulse stretching. However, the latter
results from substantial growth of the group-delay dispersion (GDD)
(quadratically with energy \cite{agrawal}). As a result, the energy scaling
requires a huge negative GDD, the obtained chirp-free soliton has a large
width, and it is not compressible linearly.

In the PDR \cite{rudolph,proctor}, the pulse is stretched and its
peak power is reduced due to large chirp \cite{haus,kalash}. The
chirp compensates the spectrum narrowing with energy and the pulse
becomes to be compressible linearly down to $T\approx 2/\Delta $,
where $\Delta $ is the spectrum half-width. The issue is that the
chirped solitary pulse (CSP) is a dissipative soliton, that is it
develops in a dissipative nonlinear system and, as a result, there
is no a uniform description of its properties and dynamics, because
the underlying nonlinear equation (so-called, nonlinear
complex Ginzburg-Landau equation, CGLE) is not integrable \cite%
{akhmed,akhmed2}.

In this work, I propose the approximate method of integration of the
generalized nonlinear CGLE and demonstrate that the CSP is its solitary
pulse solution with reduced dimension (2 or 3) of the parametric space. As a
result, the CSP characteristics become easily traceable on the
two-dimensional diagram (\textquotedblleft master diagram\textquotedblright
). Comparison of the PDR parameters demonstrates that the CSPs formed in the
ANDi fiber oscillator and in the CPO: i) lie within the distinct sectors of
the unified master diagram, ii) belong to the distinct branches of solution,
and iii) vary with parameters in different ways. Comparison of the models
based on the different versions of the master equation is carried out. The
phenomenon of concave spectrum is attributed to the quintic self-phase
modulation.

\section{Dissipative soliton of nonlinear CGLE}

The nonlinear CGLE is the generalized form of the master mode-locking
equation \cite{kaertner,akhmed,akhmed3} and provides an adequate description
of mode-locked oscillators (both fiber and solid-state). Its soliton-like
solutions (or dissipative solitons) pattern the laser pulses. Such an
approach is well-grounded if i) $T\gg 2\pi /\omega _{0}$ ($\omega _{0}$ is
the carrier frequency of laser field) and ii) relative variation of laser
field during one cavity round-trip is small.

Let $u\left( z,t\right) $ be a slowly-varying field amplitude, $z$ be a
propagation coordinate normalized to the cavity period, $t$ \ be a local
time. The generalized CGLE is
\begin{equation}
u_{z}=-\sigma u+\left( \alpha +i\beta \right) u_{tt}-iu\left( \gamma
\left\vert u\right\vert ^{2}+\chi \left\vert u\right\vert ^{4}\right) \text{
}+uf\left( \left\vert u\right\vert ^{2}\right) ,  \label{eqn1}
\end{equation}%
where $\left\vert u\right\vert ^{2}$ is the instant power, $\sigma $ is the
saturated net-loss, and $\alpha $ is the squared inverse spectral bandwidth
of oscillator (as a rule, it is defined by gain bandwidth). Parameter $\beta
$ is the net-GDD coefficient; $\gamma $ is the self-phase modulation (SPM)
coefficient, and $\chi $ describes a high-order correction to it (i.e. the
quintic SPM). Function $f\left( \left\vert u\right\vert ^{2}\right) $ models
the self-amplitude modulation (SAM) in an oscillator and its form depends on
the mode-locking mechanism.

The CSP develops in the PDR under combined action of two mechanisms:
the pure phase and dissipative ones. The first one results from a
balance of phase contributions from the pulse envelope $\beta
u_{tt}$ and the time-dependent phase $-\beta u\left( \phi
_{t}\right) ^{2}$. Such a balance is provided by some value of pulse
chirp. However, a sole phase balance is not sufficient as the pulse
spreads. The spreading can be compensated by spectral filtering.
Since the chirp causes the frequency deviation at pulse front and
tale, the filter cuts off the higher- and lower-frequency wings of
the pulse and, thereby, shortens it\cite{haus,proctor}.

The partial exact CSP solution of Eq. (\ref{eqn1}) is known for $f\left(
\left\vert u\right\vert ^{2}\right) =\kappa \left( 1-\varsigma \left\vert
u\right\vert ^{2}\right) \left\vert u\right\vert ^{2}$ (cubic-quintic
nonlinear CGLE, for overview see \cite{akhmed,akhmed2}). In this work, the
approximate method of integration of Eq. (\ref{eqn1}) in a general form will
be proposed. The underlying approximations are

\begin{conjecture}
$T\gg \sqrt{\beta }$, that is the adiabatic approximation;
\end{conjecture}

\begin{conjecture}
$\beta \gg \alpha $, that is the GDD prevails over the spectral dissipation.
\end{conjecture}

The first conjecture is valid for both solid-state and fiber oscillators
operating in the PDR because the pulse is strongly stretched in the regime
under consideration. The second conjecture is valid for both broadband
solid-state (i.e. Ti:Sapphire \cite{kalash} and Cr:YAG \cite{kalash2}) and
fiber oscillators\cite{wise2}. Thin-disk oscillators based on the narrowband
active media\cite{keller} can approach the limit of $\beta \approx \alpha $
and this issue will be addressed in Subsection 2.4.

\subsection{Cubic nonlinear CGLE}

The simplest version of (\ref{eqn1}) corresponds to $\chi =0$ and
the SAM function is\cite{haus}

\begin{equation}
f\left( \left\vert u\right\vert ^{2}\right) =\kappa \left\vert u\right\vert
^{2}.  \label{sh1}
\end{equation}

\noindent Such a function approximates nonlinearity of low-energy
solid-state and fiber oscillators. The $\kappa $-parameter describes a
nonlinear gain due to loss saturation. Eqs. (\ref{eqn1},\ref{sh1}) lead to a
dissipative generalization of the nonlinear Schr\"{o}dinger equation.

Let's make the traveling wave reduction of Eqs. (\ref{eqn1},\ref{sh1}) by
means of ansatz

\begin{equation}
u\left( z,t\right) =\sqrt{P\left( t\right) }\exp \left[ i\phi \left(
t\right) -iqz\right] ,  \label{trwave}
\end{equation}

\noindent where $P$ is the $z$-independent instant power, $\phi \left(
t\right) $ is the time-dependent phase, and $q$ is the phase due to slip of
the carrier phase with respect to the envelope. In contrast to the tradition
approach\cite{akhmed}, we do not impose hereinafter any restriction on the
time-dependence of phase. Substitution of (\ref{trwave}) in (\ref{eqn1})
supplemented with (\ref{sh1}) as well as taking into account the
approximations under consideration and the obvious restrictions $P>0$ and $%
\phi _{tt}<\infty$ lead to

\begin{eqnarray}
\gamma P\left( t\right) &=&\beta \Delta ^{2}\left( 1-\tanh ^{2}\left[
t\kappa \left( 1+c\right) \Delta /3\gamma \right] \right)  \notag \\
\alpha \Delta ^{2} &=&\frac{3\sigma c}{2-c},\ \gamma P\left( 0\right) =\beta
\Delta ^{2},  \label{sh2}
\end{eqnarray}

\noindent where $\Delta ^{2}\equiv q/\beta $, $P\left( 0\right) $ is the CSP
peak power, and the control parameter $c\equiv \alpha \gamma /\beta \kappa $%
. One can see, that, on conditions that the appropriate normalizations are
used, the CSP is two-parametric and depends on only $\sigma $ and $c$.

Since a pulse is strongly chirped in the PDR, one may treat $\phi $
as a rapidly varying function and apply the method of stationary
phase to the Fourier image of $u$\cite{kalash3}. As a result, the
spectral power is

\begin{equation}
p\left( \omega \right) \equiv |e\left( \omega \right) |^{2}\simeq \frac{6\pi
\beta }{\left( 1+c\right) \kappa }H\left( \Delta ^{2}-\omega ^{2}\right) ,
\label{sh3}
\end{equation}

\noindent where $e\left( \omega \right) \equiv \tciFourier \left[ u\right]
=\int dt\sqrt{P\left( t\right) }\exp \left[ i\phi \left( t\right) -i\omega t%
\right] $ and $H\left( x\right) $ is the Heaviside function. That is the
spectrum is flat-top and truncated at $\pm \Delta $. The latter parameter
plays a role of the spectrum half-width. From Eq. (\ref{sh3}) the CSP energy
is

\begin{equation}
E\equiv \int_{-\infty }^{\infty }Pdt=\int_{-\Delta }^{\Delta }\frac{d\omega
}{2\pi }p=\frac{6\beta \Delta }{\left( 1+c\right) \kappa }.  \label{sh4}
\end{equation}

Important features of (\ref{sh2}) are i) $0<c<2$, ii) there is no
physically nontrivial limit $\sigma =0$, and iii) spectral width
increases with $\sigma $ and $c$. Hereinafter, the solutions with
the properties of ii) and iii) will be termed as the Schr\"{o}dinger
(or negative) branch of CSP.

\subsection{Cubic-quintic nonlinear CGLE ($\protect\chi =0$)}

Let's consider the case with negligible higher-order SPM and with
the SAM in the form of\cite{kalash}

\begin{equation}
f\left( \left\vert u\right\vert ^{2}\right) =\kappa \left( 1-\varsigma
\left\vert u\right\vert ^{2}\right) \left\vert u\right\vert ^{2}.
\label{eqn2}
\end{equation}

\noindent Here the $\varsigma $-parameter corresponds to the SAM
saturation. Such a form of SAM can be attributed to the Kerr-lens
mode locking or the mode-locking due to polarization modulator. The
former technique uses power-dependence of the laser beam size due to
self-focusing inside a nonlinear medium. As a result, the
overlapping between the laser and pump beams becomes
power-dependent, as well. This leads to the SAM in the form under
consideration. For the mode-locking technique utilizing the
power-dependent polarization, Eq. (\ref{eqn2}) can be considered as
a low-order in $P$ approximation of the trigonometric SAM function.
The SPM saturation (i.e. the $\chi $-term) can be omitted if the
beam confocal length inside an active crystal is much less than the
crystal length (in a solid-state oscillator) or the mode is strongly
confined (in a fiber oscillator). Also, $\chi =0$ for an airless
thin-disk oscillator.

Substitution of (\ref{trwave}) in (\ref{eqn1}) supplemented with (\ref{eqn2}%
) and taking into account the approximation under consideration result in%
\cite{kalash}

\begin{eqnarray}
\gamma P &=&q-\beta \Omega ^{2},  \label{eqn4} \\
\beta \left( \Omega _{t}+\frac{\Omega }{P}P_{t}\right) &=&\kappa P\left(
1-\varsigma P\right) -\sigma -\alpha \Omega ^{2},  \notag
\end{eqnarray}

\noindent where $\Omega \equiv \phi _{t}$ is the frequency deviation from the carrier frequency $%
\omega _{0}.$

The restrictions $P>0$ and $\Omega _{t}<\infty $ allow obtaining

\begin{eqnarray}
\gamma P\left( 0\right) &=&\beta \Delta ^{2}=\frac{3\gamma }{4\varsigma }%
\left( 1-c/2\pm \sqrt{\left( 1-c/2\right) ^{2}-4\sigma \varsigma /\kappa }%
\right) ,  \notag \\
\Omega _{t} &=&\frac{\beta \varsigma \kappa }{3\gamma ^{2}}\left( \Delta
^{2}-\Omega ^{2}\right) \left( \Omega ^{2}+\Xi ^{2}\right) ,  \label{eqn5} \\
\beta \Xi ^{2} &=&\frac{\gamma }{\varsigma }\left( 1+c\right) -\frac{5}{3}%
\gamma P\left( 0\right) .  \notag
\end{eqnarray}

\noindent Integration of second Eq. (\ref{eqn5}) in the combination with
first Eq. (\ref{eqn4}) gives the implicit expression for CSP profile\cite%
{kalash3}. To define the spectral shape of CSP and its energy, one may use
the approach described in previous Subsection. The spectral power is

\begin{equation}
p\left( \omega \right) \simeq \frac{6\pi \gamma }{\varsigma \kappa }\frac{%
H\left( \Delta ^{2}-\omega ^{2}\right) }{\Xi ^{2}+\omega ^{2}}.  \label{eqn6}
\end{equation}

\noindent That is the CSP spectrum has the truncated Lorentz profile, where $%
\Delta $ plays a role of the spectral half-width (if $\Xi >\Delta $,
otherwise the spectral width is defined by $\Xi $).

Eq. (\ref{eqn6}) allows obtaining the CSP energy

\begin{equation}
E=\frac{6\gamma }{\varsigma \kappa \Xi }\arctan \left( \frac{\Delta }{\Xi }%
\right) .  \label{eqn7}
\end{equation}

Hence, Eqs. (\ref{eqn5},\ref{eqn6},\ref{eqn7}) define the CSP completely.
One can see, that the CSP parameters depend on only two control parameters: $%
c\equiv \alpha \gamma /\beta \kappa $ and $a\equiv \sigma \varsigma /\kappa $%
. Since the saturated net-gain is energy-dependent due to gain saturation,
one may choose $c$ and $E^{\ast }$ as two control parameter, where $E^{\ast
} $ is the energy of steady-state (continuous-wave, CW) solution of
linearized Eq. (\ref{eqn1}).

Thus, the CSP characteristics can be mapped on two-dimensional diagram
(so-called \textquotedblleft master diagram\textquotedblright ), that makes
theirs easily traceable (see below).

\subsection{Cubic-quintic nonlinear CGLE ($\protect\chi \neq 0$)}

Eqs. (\ref{eqn1},\ref{eqn2}) with $\chi \neq 0$ describe an
oscillator operating under condition, that the beam is not confined
strongly in a nonlinear element and its size variation affects the
effective SPM. For instance, that can be a solid-state oscillator
with the beam confocal length approaching or exceeding the crystal
length; a high-energy oscillator with nonlinear plate providing the
SPM; a fiber oscillator with no strong mode confinement, etc.

The method sketched in previous Subsections leads to

\begin{eqnarray}
P &=&\frac{b}{2}\left[ \Psi -1\right] ,  \label{eqn8} \\
\Delta ^{2} &=&\frac{c}{16\left( 1+\frac{c}{b}\right) }\left[ \frac{2\left(
3+\frac{4}{b}+\frac{c}{b}\right) }{1+\frac{c}{b}}\left( 2+\frac{c}{2}+\frac{%
3b}{2}\pm \sqrt{\left( 2-c\right) ^{2}-16a\left( 1+\frac{c}{b}\right) }%
\right) -3c-9b-\frac{32a}{b}-12\right] ,  \notag \\
\Omega _{t} &=&-\frac{c\Psi \left( 1-\Psi \right) \left( a+\Omega ^{2}+\frac{%
b^{2}}{4}\left( 1-\Psi \right) \left( 1-\Psi +\frac{2}{b}\right) \right) }{%
\Psi \left( 1-\Psi \right) +4\Omega ^{2}/cb},  \notag
\end{eqnarray}

\noindent where $\Psi \left( t\right) \equiv
\sqrt{1+\frac{4}{cb}\left( \Delta ^{2}-\Omega ^{2}\left( t\right)
\right) }$, $b \equiv \varsigma \gamma /\chi$ and the following
normalizations are used: $t^{^{\prime }}=t\kappa \sqrt{\kappa
/\alpha \varsigma }/\varsigma $, $P^{^{\prime }}=\varsigma P$,
$\Omega ^{^{\prime }2}=\Omega ^{2}\alpha \varsigma /\kappa $,
$\Delta ^{^{\prime }2}=\Delta
^{2}\alpha \varsigma /\kappa $ (the primes will be omitted below). Eqs. (\ref%
{eqn8}) represent only solutions, which tend to those of Eqs. (\ref{eqn5})
when $\chi \longrightarrow 0$.

Expression for the dimensionless spectral power is

\begin{equation}
p\left( \omega \right) \simeq \frac{\pi \left( A-1\right) \left(
\left( A-1\right) cb+4\left( 2\omega ^{2}-\Delta ^{2}\right) \right)
H\left( \Delta ^{2}-\omega ^{2}\right) }{cA\left( \left( A-1\right)
\left( c\left( a+b+b^{2}+\omega ^{2}\right) +b\left( \Delta
^{2}-\omega ^{2}\right) \right) -2\left( b+1\right) \left( \Delta
^{2}-\omega ^{2}\right) \right) }, \label{eq9}
\end{equation}

\noindent where $A = \sqrt {1 + \frac{{4\left( {\Delta ^2  - \omega
^2 } \right)}} {{cb}}}$. One can see, that the CSP becomes
three-parametric due to an appearance of non-zero $\chi $ (i.e.
$b\neq \infty $).

The CSPs under consideration subdivide into two classes accordingly
two signs in Eqs. (\ref{eqn5},\ref{eqn8}): i) positive branch and
ii) negative (or Schr\"{o}dinger) branch. The negative branch can be
transformed into the soliton-like solution of dissipative nonlinear
Schr\"{o}dinger equation, when the quintic nonlinear terms tend to
zero (see Subsection 2.1 and \cite{malomed}). The positive branch,
possessing physically nontrivial limit $a=0$, does not allow such a
transformation. The additional characteristics of these two branches
will be described in detail below.

\begin{figure}
 \includegraphics[width=10cm]{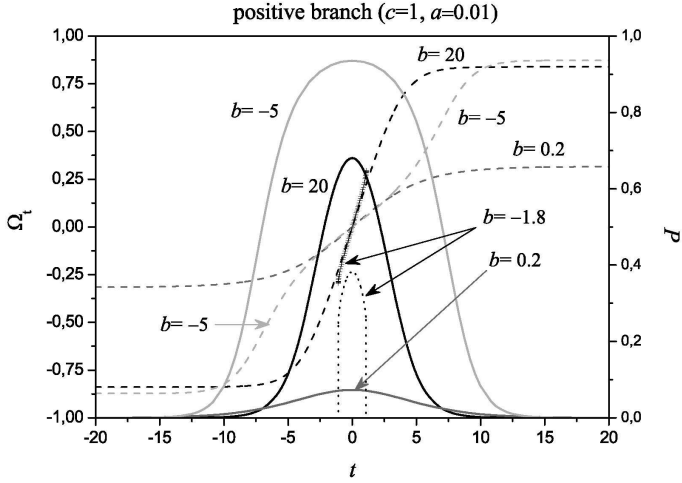}
  \caption{Positive branch: the CSP profiles $P(t)$ (solid and dotted curves) and the corresponding chirps $\Omega
_{t}(t)$ (dashed curves and crosses) for the different values of
quintic term $b$. $a=0.01$, $c=1$.\label{f1}}
\end{figure}

\begin{figure}
  \includegraphics[width=10cm]{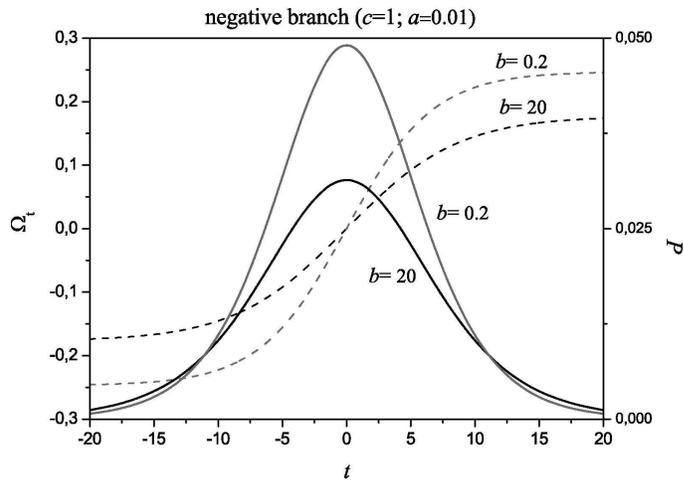}
  \caption{Negative branch: the CSP profiles $%
P(t)$ (solid curves) and the corresponding chirps $\Omega _{t}(t)$
(dashed curves) for the different values of quintic term $b$.
$a=0.01$, $c=1$.\label{f2}}
\end{figure}

Let's consider the CSP profiles obtained from integration of (\ref{eqn8}).
Figs. \ref{f1} and \ref{f2} show the profiles of the positive and negative
branches, respectively. The $b$-parameter is scalable and zero quintic SPM
corresponds to $b\longrightarrow \pm \infty $. When $b>0$, nonlinear phase
shift increases with power. Such a sign of $b$ corresponds, for instance, to
a Kerr-lens mode-locked oscillator. One can see, that the peak power and,
correspondingly, the chirp increases (decreases) with $b$ for positive
(negative) branch. When $b<0$ (saturable SPM), both flat-top (gray solid
curve, Fig. \ref{f1}) and parabolic (dotted curve, Fig. \ref{f1})\ profiles
appear.

\begin{figure}
  \includegraphics[width=10cm]{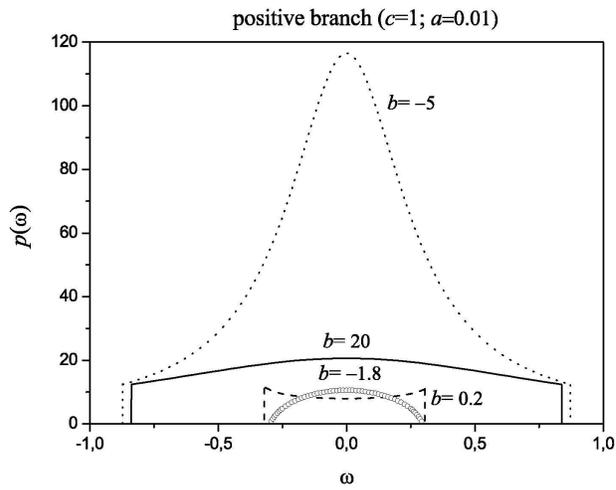}
  \caption{Positive branch: the CSP spectra
for the different values of quintic term $b$. $a=0.01$,
$c=1$.\label{f3}}
\end{figure}

As it has been pointed (see Eqs. (\ref{sh3},\ref{eqn6},\ref{eq9})),
the CSP spectra are truncated at some frequency $\pm \Delta $. The
spectral profiles of positive branch are shown in Fig. \ref{f3}. One
can see, that there exist next spectral types: i)
parabolic-top\cite{kalash,wise2} (solid curve and circles; the last
corresponds to large contribution of saturable SPM), ii)
finger-like\cite{kalash,wise2} (dotted curve; that is the truncated
Lorentz profile), and iii) concave\cite{wise2,wise3,kalash4} (dashed
curve). In contrast to the model presented in Ref.\cite{wise3}, our
model predicts that the concave spectra, which are widely presented
in the fiber oscillators, are stable. The last conclusion means that
such spectra exist for the saturable SAM (i.e. for $\varsigma >0$).
The cause of concave spectra is the positive quintic SPM (i.e. the
SPM growing with $P$).

\begin{figure}
  \includegraphics[width=10cm]{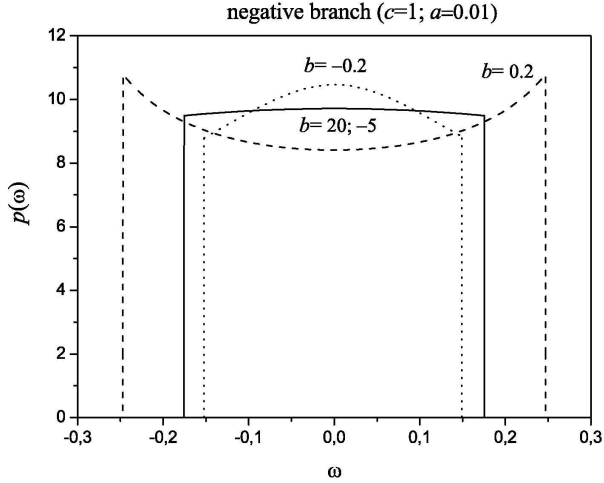}
  \caption{Negative branch: te CSP
spectra for the different values of quintic term $b$. $a=0.01$, $c=1$. The profiles for $%
b=20$ and -5 coincide.\label{f4}}
\end{figure}

Spectra of negative branch are shown in Fig. \ref{f4}. The spectra
are typically narrower than those of positive branch. Therefore, the
finger-like profiles disappear and the typical shapes are i)
parabolic- or flat-top, and ii) concave. The latter shape correspond
to wider spectra formed in the presence of positive quintic SPM
(i.e. the SPM growing with $P$).

Dependence of the normalized spectral half-width $\Delta $ on the normalized
saturated net-loss parameter $a$ for the different $b$ is shown in Fig. \ref%
{f5}. When $b>0$ or $b<-2$, the positive branch CSP has a wider spectrum
than the negative branch one. Under these conditions, the spectrum width of
positive (negative) branch decreases (increases) with the $a$-growth. The
behavior of negative branch corresponds to that of CSP in the dissipative
nonlinear Schr\"{o}dinger equation (see Subsection 2.1).

The CSP exists within confined region of $a$, which narrows with the
decreasing positive $b$. When $0>b\geq -2$, the positive branch exists only
for nonzero $a$. This means that only negative-branch CSP can spontaneously
develop in an oscillator with $0>b\geq -2$.

\begin{figure}
  \includegraphics[width=10cm]{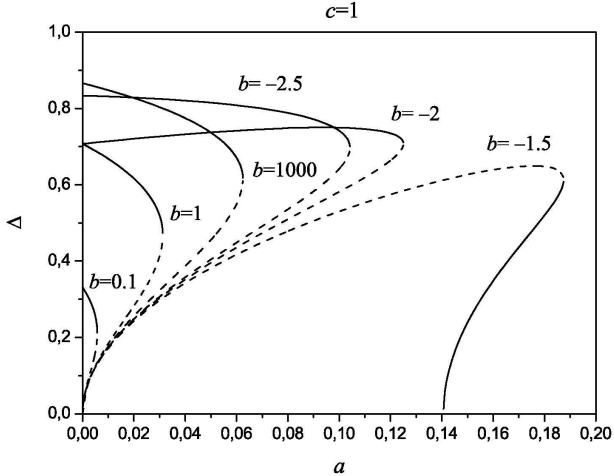}
  \caption{The normalized spectral half-width
in dependence on the saturated net-loss parameter $a$ for the
different values of quintic SPM. $c=1$. Positive branch - solid
curves; negative branch - dashed curves. \label{f5}}
\end{figure}

\begin{figure}
  \includegraphics[width=10cm]{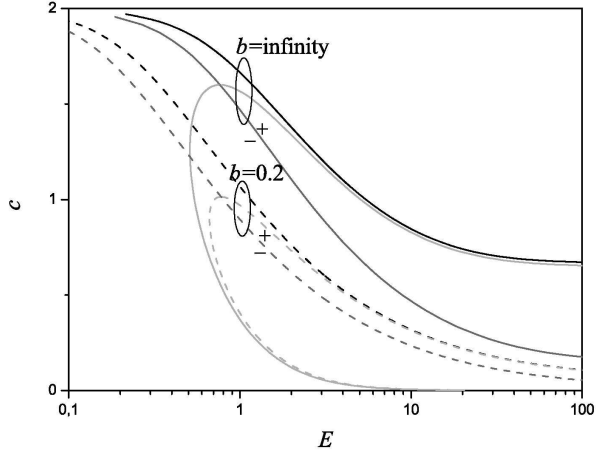}
  \caption{Master diagram: $%
b\longrightarrow \infty $ (solid curves), $b=0.2$ (dashed curves).
Black curves are the stability thresholds ($a=0$) , gray curves are
the border between the positive ($+$) and negative ($-$) branches,
light gray curves are the positive and negative branches for
$a=0.01$. \label{f6}}
\end{figure}

\begin{figure}
  \includegraphics[width=10cm]{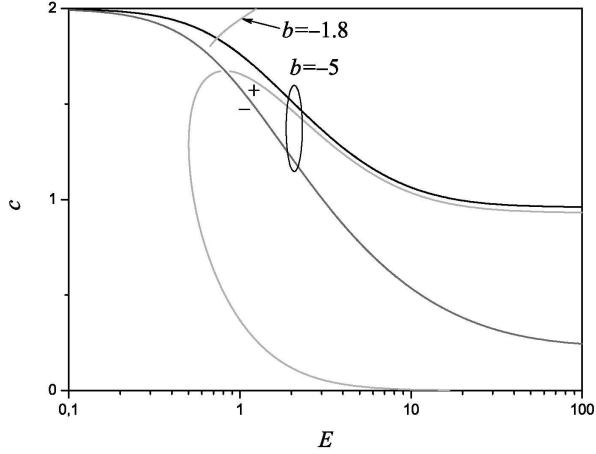}
  \caption{Master diagram: $b=-5$ (solid
curves), $b=-1.8$ (labeled gray curve). Black curve is the stability
threshold ($a=0$) , gray curve is the border between the positive
($+$) and negative ($-$) branches, light gray curves are the
positive and negative branches for $a=0.01$. Gray curve $b=-1.8$
corresponds to the negative branch. \label{f7}}
\end{figure}

Figs. \ref{f6},\ref{f7} show the master diagrams, which are the
two-dimensional slices of three-dimensional parametric space of CSP
(for $\chi \neq 0$). The dimensionless parameter $E$ equals to the
dimensional energy $E$ multiplied by $\left( \kappa /\gamma \right)
\sqrt{\kappa \varsigma /\alpha }$ and can be easily related to
$E^{\ast }$ by means of expansion of $\sigma $ in the vicinity of
laser threshold\cite{kalash}: $\sigma \approx \delta \left(
E/E^{\ast }-1\right) $ ($ \delta  \equiv \left. {{{d\sigma }
\mathord{\left/
 {\vphantom {{d\sigma } {dE}}} \right.
 \kern-\nulldelimiterspace} {dE}}} \right|_{E = E^* }
$). The
black curves (the solid ones in both Figs. and the black dashed one in Fig. %
\ref{f6}) correspond to the stability threshold against the
continuum excitation, that is $a=0$ along this curve. The stable CSP
exists below these curves (i.e. the curves correspond to the maximum
values of $c$ for a given $E$). One can see, that the stability
threshold $c$ decreases with $E$. Physically, that
means, for instance, the growth of GDD required for the CSP stabilization ($%
c\propto 1/\beta $). It is important, that there is an asymptotic value of $c
$ (i.e. an extra-growth of $E$ does not change the threshold value of $c$
substantially). The contribution of positive (negative) quintic SPM narrows
(broadens) the stability region. Also, the region, where the positive branch
exists, narrows for the growing positive $\chi $ (i.e. with the decrease of
positive $b$). It should be noted, that the positive branch disappears for
small $a$, when $0>b\geq -2$.

The master diagrams demonstrate main difference between the CSP
branches. When $b>0$ is not too small, the asymptotic behavior of
isogains (i.e. the curves of constant $a$) with the $E$-growth
demonstrates that the CSP energy is scalable. That is $E\propto
E^{\ast }$ and the proportionality coefficient is weakly dependent
on the parameters of (\ref{eqn1}). In this sense, the negative
(Schr\"{o}dinger) branch is not energy-scalable, because the energy
depends on $c$ weakly and the change of $E$ requires a substantial
change of $c$ (e.g. a substantial GDD growth as $c\propto 1/\beta
$).

\subsection{Generalized nonlinear CGLE}

In this subsection the SAM corresponding to a perfectly saturable absorber%
\cite{wise2,kalash4} will be considered. This type of SAM represents
a semiconductor saturable absorber mirror (SESAM), which are
extensively used in CPO
oscillators\cite{keller,naumov,morgner2,neuhaus}. If the CSP width
(usually, few picoseconds) excesses the SESAM relaxation time
$T_{r}$ (hundreds of femtoseconds), the SAM function can be written
in the form

\begin{equation}
f\left( \left\vert u\right\vert ^{2}\right) =\frac{\kappa \left\vert
u\right\vert ^{2}}{1+\varsigma \left\vert u\right\vert ^{2}},  \label{eqn10}
\end{equation}

\noindent where $\kappa =\mu \varsigma $, $\mu $ is the modulation depth, $%
\varsigma =T_{r}/E_{s}S$ is the inverse saturation power ($E_{s}$ is
the SESAM saturation energy fluency, $S$ is the beam area on SESAM).
We assume below that $\chi=$0.

Eqs. (\ref{eqn1},\ref{eqn10}) can be reduced by the above described method to

\begin{eqnarray}
\varsigma P\left( 0\right) &=&\frac{\alpha \Delta ^{2}}{\mu c}=\frac{3}{4c}%
\left( 2\left( 1-a\right) -c\pm \sqrt{\Upsilon }\right) ,  \label{eqn11} \\
\Omega _{t} &=&\frac{\alpha }{3\beta }\frac{\left( \Delta ^{2}-\Omega
^{2}\right) \left( \Xi ^{2}-\Omega ^{2}\right) }{\Delta ^{2}-\Omega
^{2}+\gamma /\varsigma \beta },  \notag \\
\frac{\alpha \Xi ^{2}}{\mu } &=&\frac{2\alpha }{3\mu }\Delta ^{2}+1-a+c.
\notag
\end{eqnarray}

\noindent Here $a\equiv \sigma /\mu $, $c\equiv \alpha \gamma /\beta \kappa $%
, $\Upsilon \equiv \left( 2-c\right) ^{2}-4a\left( 2-a+c\right) $. The
spectral profile (truncated parabolic- or flat-top) and the CSP energy are

\begin{eqnarray}
p\left( \omega \right) &\approx &\frac{6\pi \beta ^{2}}{\alpha \gamma }\frac{%
\Delta ^{2}-\omega ^{2}+\gamma /\varsigma \beta }{\Xi ^{2}-\omega ^{2}}%
H\left( \Delta ^{2}-\omega ^{2}\right) ,  \label{eqn12} \\
E &\approx &\frac{6\beta ^{2}\Delta }{\alpha \gamma }\left[ 1-\frac{\left(
\Xi ^{2}-\Delta ^{2}-\gamma /\beta \varsigma \right) ~arc\tanh \left( \frac{%
\Delta }{\Xi }\right) }{\Delta \Xi }\right] .  \label{eqn13}
\end{eqnarray}

Using the normalizations from previous Subsection reduces Eq. (\ref{eqn13})
to

\begin{equation}
E=\frac{6\Delta }{c^{2}}\left[ 1-\frac{\left( \Xi ^{2}-\Delta ^{2}-c\right)
~arc\tanh \left( \frac{\Delta }{\Xi }\right) }{\Delta \Xi }\right] ,
\label{eqn14}
\end{equation}

\noindent where primes for the normalized values are omitted.

\begin{figure}
  \includegraphics[width=10cm]{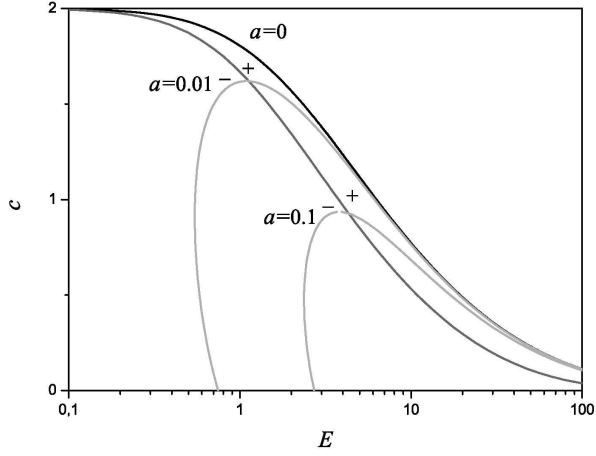}
  \caption{Master diagram of generalized
nonlinear CGLE. Black curve is the stability threshold ($a=0$) ,
gray curve is the border between the positive ($+$) and negative
($-$) branches, light gray curves are the positive and negative
branches for $a=0.01$ and $0.1$. \label{f8}}
\end{figure}

The master diagram following from Eq. (\ref{eqn14}) is presented in Fig. \ref%
{f8}. The structure of diagram is similar to that described in previous
Subsection, but there are two important differences: i) $c$-parameter
asymptotically tends to zero with growth of $E$, ii) negative branch is not
energy-scalable within all range of existence. The latter conclusion is
obvious from behavior of isogains in Fig. \ref{f8} as one can see that the $c
$-scaling weakly affects $E$. Thus, the energy remains almost constant along
this isogain when the GDD scales ($c\propto 1/\beta $). However, the energy
scaling for the negative branch can be provided by a simultaneous growth of $%
\alpha $ and $\beta $ so that $c$ remains constant (see also the
normalization for $E$).

\begin{figure}
  \includegraphics[width=10cm]{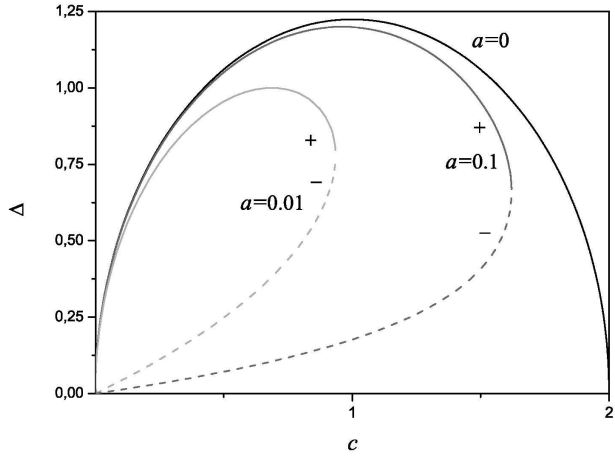}
  \caption{Normalized spectral half-width
in dependence on $c$ for the different values of $a$. Positive
branch - solid curves; negative branch - dashed curves. \label{f9}}
\end{figure}

For the negative branch, the spectrum narrows with the $c$-decrease
due to growth of the GDD contribution (Fig. \ref{f9}), which
stretches the pulse when the energy remains almost constant. When
$E$ changes weakly along the
isogain corresponding to the negative branch, the spectrum broadens with $%
\alpha $ (Fig. \ref{f9}; $c\propto \alpha $). The explanation is that the
growth of spectral filtering enhances the cutoff of red (blue)-shifted
spectral components located on the pulse front (tail). The growth of cutoff
shortens the CSP and, for a fixed energy, $P\left( 0\right) $ increases.
Since $\Delta ^{2}\propto P\left( 0\right) $, the spectrum broadens. For the
positive branch, the spectrum initially broadens with $c$-decrease (Fig. \ref%
{f9}). This can be explained as a result of $E$-increase, which is necessary
for keeping in the isogain (Fig. \ref{f8}). That is, the SPM contribution
increases and the spectrum broadens. However, further decrease of $c$
narrows the spectrum due to growth of the CSP width. The latter results from
either GDD growth ($c\propto 1/\beta $) or suppression of spectral cutoff on
the pulse wings due to $\alpha $-decrease ($c\propto \alpha $).

The CSP spectral width behaves similarly for the cubic-quintic
nonlinear CGLE with $\chi =0$. It should be noted additionally, that
the positive branch remains energy-scalable for the SAM under
consideration, but there is a need in the decrease of $c$ to provide
such a scaling (Fig. \ref{f8}).

One has to emphasize that the master diagrams cover both solid-state and
fiber oscillators. In a fiber oscillator, the GDD value is larger in
comparison with that in a solid-state CPO, but the spectral filter bandwith (%
$\sim $25 nm) is small. As a result, the excess of the ratio $\beta /\alpha $
over that for a Ti:sapphire CPO is only tenfold\cite{kalash4}.
Simultaneously, an excess of the ratio $\gamma /\kappa $ over that for a
Ti:sapphire CPO is tenfold as well. As a result, the $c$-parameter is $%
\simeq 1$ and, dynamically, there is no substantial distinction in kind
between the ANDi fiber and the solid-state CPOs. One difference is that a
larger GDD and a comparatively smaller $E$ shift the operational point of a
fiber oscillator into the negative branch region. While the operational
point of a solid-state oscillator belongs to the positive branch. As a
result, their scaling properties are different.

In the case of narrow-band thin disk solid-state oscillator, the assumption $%
\beta \gg \alpha $ is violated. The result is the smoothed spectrum
edges. Such a smoothing increases with the decrease of SPM (e.g.,
when a resonator becomes airless). Nevertheless, the numerical
simulations demonstrate that
the analytical model describes the spectral width of CSP adequately even for $%
\alpha \approx \beta $. However, the case of $\alpha >\beta $ is
beyond the bounds of the theory under consideration.

\section{Conclusions}

The completely analytical theory of CPOs has been developed. It has
been found, that such oscillators (both solid-state and fiber) can
be modeled by the nonlinear CGLE, which has been integrated
approximately on the basis of the proposed method. As a result, one
may easily trace the properties of CSPs, because the CSP has two-
(maximum three-)dimensional representation (so-called, master
diagram). Four types of SAM have been analyzed on the common basis.
These SAMs are inherent in both solid-state and fiber CPOs
mode-locked by either self-focusing, or SESAM, or polarization
modulator. The CSPs have been found to be subdivided into two
classes with different scaling properties. This properties have been
traced and described within the whole parametric space.

The mathematical apparatus is presented in detail in
http://info.tuwien.ac.at/kalashnikov/NCGLE1.html\ and
http://info.tuwien.ac.at/kalashnikov/genNCGLE.html.

\acknowledgments

The work was supported by Austrian Fonds zur F\"{o}rderung der
wissenschaftlichen Forschung (project P20293).

\bibliographystyle{spiebib}
\bibliography{report}

\end{document}